\documentclass[preprint]{aastex}
\usepackage{mkfig}

\begin{document}

\title{\bf Hydrogen Lyman-$\alpha$ Absorption Predictions by Boltzmann Models
  of the Heliosphere}

\author{Brian E. Wood\altaffilmark{1}}
\affil{JILA, University of Colorado and NIST, Boulder, CO 80309-0440.}
\email{woodb@marmot.colorado.edu}

\and

\author{Hans-Reinhard M\"{u}ller, Gary P. Zank}
\affil{Bartol Research Institute, University of Delaware, Newark, DE 19716.}
\email{mueller@bartol.udel.edu, zank@bartol.udel.edu}

\altaffiltext{1}{Visiting Scientist, Bartol Research Institute}

\begin{abstract}

     We use self-consistent kinetic/hydrodynamic models of the heliosphere
to predict H~I Lyman-$\alpha$ absorption profiles for various lines of sight
through the heliosphere.  These results are compared with Lyman-$\alpha$
absorption lines of six nearby stars observed by the {\em Hubble Space
Telescope}.  The directions of these lines of sight range from
nearly upwind (36~Oph) to nearly downwind ($\epsilon$~Eri).  Only three of
the Lyman-$\alpha$ spectra (36~Oph, $\alpha$~Cen, and Sirius) actually show
evidence for the presence of heliospheric absorption, which is blended with
the ubiquitous interstellar absorption, but the other three spectra still
provide useful upper limits for the amount of heliospheric absorption for
those lines of sight.  Most of our models use a Boltzmann particle code for
the neutrals, allowing us to estimate neutral velocity distributions
throughout the heliosphere, from which we compute model Lyman-$\alpha$
absorption profiles.  In comparing these models with the data, we find they
predict too much absorption in sidewind and downwind directions, especially
when higher Mach numbers are assumed for the interstellar wind.
Models created assuming different values of the interstellar temperature and
proton density fail to improve the agreement.  Somewhat surprisingly,
a model that uses a multi-fluid treatment of the neutrals rather than
the Boltzmann particle code is more consistent with the data, and we
speculate as to why this may be the case.

\end{abstract}

\keywords{hydrodynamics --- shock waves --- solar wind --- ultraviolet: ISM}

\section{Introduction}

     The basic heliospheric structure created by the interaction between
the fully ionized solar wind and the partially ionized local interstellar
medium (LISM) can to first order be modeled by considering plasma
interactions alone, and therefore many models both new and old have been
presented that ignore the neutral LISM particles entirely
\citep[e.g.,][]{teh89,rss94,cw98}.
However, the charge exchange mechanism, whereby an interstellar neutral loses
its electron to a proton, allows the neutrals to take part in the
solar-wind/LISM interaction in many important ways
\citep[see review by][]{gpz99}.

     For example, many interstellar neutrals that penetrate the heliopause
are ionized by charge exchange with solar wind protons and are then picked up
by the outflowing solar wind.  These ``pick-up ions'' dominate the internal
energy of the solar wind beyond about 10 AU \citep{lfb94,gg94},
and the acceleration of some of these ions at the termination shock is
believed to be the origin of anomalous cosmic rays \citep{laf74,bk95}.

     Heliospheric models that treat the neutral gas and plasma in a
self-consistent manner predict somewhat different properties than plasma-only
models, such as shorter distances to the termination shock, heliopause, and
bow shock \citep{vbb93,vbb95,hlp95,gpz96}.
These models also suggest that neutral hydrogen in the
heliosphere should be very hot, with temperatures of order 20,000--40,000 K.
The importance of this prediction is that this high temperature gas should
produce H~I Lyman-$\alpha$ absorption broad enough to be separable from the
interstellar absorption observed toward nearby stars, meaning hydrogen in
the outer heliosphere can potentially be directly detectable.  This
absorption has in fact been observed using observations by the {\em Hubble
Space Telescope} (HST).  \citet{jll96} detected excess Lyman-$\alpha$
absorption in HST spectra of the nearby stars $\alpha$~Cen A and B.  They and
\citet{kgg97} both demonstrated that the properties of this excess
absorption are consistent with a heliospheric origin.

     In upwind directions, such as that toward $\alpha$~Cen, the heliospheric
H~I column density is dominated by compressed, heated, and decelerated
material just outside the heliopause, which constitutes the so-called
``hydrogen wall.''  This material accounts for most of the non-LISM
absorption observed toward $\alpha$~Cen.  An additional detection of
heliospheric H~I absorption only $12^{\circ}$ from the upwind direction was
provided by HST observations of 36~Oph \citep{bew00}.  For
downwind lines of sight the H~I density is much lower than in the hydrogen
wall, but the sightline through the heated heliospheric H~I is longer,
potentially allowing heliospheric Lyman-$\alpha$ absorption to be observed
downwind as well as upwind \citep{llw97}.  \citet{vvi99}
report a detection of heliospheric absorption along a downwind line of sight
toward the star Sirius.

     In addition to detections of hot H~I surrounding the Sun, there
have also been many reported detections of analogous ``astrospheric''
material surrounding other Sun-like stars \citep{bew96,ard97,bew98}.
\citet{bew98} have illustrated how these observations might potentially be
used to infer properties of the winds of these stars and their interstellar
environments.

     Models that self-consistently treat the neutrals and
plasma are essential to help interpret these observations, but such models
are a complex theoretical and computational problem.  The fundamental
difficulty is that neutrals in the heliosphere are far from equilibrium, and
their velocity distribution at a given location can be highly non-Maxwellian.
One approach is to treat the plasma as one fluid and the neutral hydrogen as
three separate fluids, one for each distinct environment in which charge
exchange occurs \citep{gpz96}.  \citet{kgg97} used such
four-fluid models to reproduce the Lyman-$\alpha$ absorption observed toward
$\alpha$~Cen.  However, the four-fluid treatment of the problem remains an
approximation and may not be entirely successful in reproducing the actual
H~I velocity distribution \citep{vbb98}, or the
expected Lyman-$\alpha$ absorption for a given line of sight.

     \citet{asl98} adopted a particle-mesh method for solving the neutral
Boltzmann equation, which should yield accurate particle distribution
functions, and \citet{hrm00} further developed this code
\citep[see also][]{hrm99}.  In this paper, we use heliospheric models
created by this Boltzmann code to predict heliospheric Lyman-$\alpha$
absorption profiles that can be compared with the HST observations.  Models
are created assuming different Mach numbers for the inflowing LISM to see
which best matches the observed amount of absorption.  For the $\alpha$~Cen
line of sight, a similar analysis has been carried out by \citet{kgg97}
but using four-fluid rather than Boltzmann models, and assuming
different LISM parameters.  We compare the absorption predicted by the models
with lines of sight observed by HST, including $\alpha$~Cen
and the aforementioned 36~Oph and Sirius lines of sight, which all show
evidence for heliospheric absorption \citep*{bew00,vvi99}.

\section{The Observations}

     By now, at least twenty or so useful lines of sight through the LISM
have been observed by HST \citep[for a partial list, see][]{jll98}.  Evidence
for heliospheric absorption has only been found for three of them, but some
of the other lines of sight with low LISM column densities might still be
useful for providing upper limits for heliospheric absorption.  Thus, in
addition to the three lines of sight that suggest heliospheric absorption
(36~Oph~A, $\alpha$~Cen, and Sirius) we also work with three other lines of
sight (31~Com, $\beta$~Cas, and $\epsilon$~Eri) that sample different
directions through the heliosphere.

     In Figure 1, we display the Lyman-$\alpha$ lines of these six stars.
The $\theta$ values shown in the figure are the angles from the upwind
direction, which range from the nearly upwind line of sight toward 36~Oph~A
($\theta=12^{\circ}$) to the nearly downwind line of sight toward
$\epsilon$~Eri ($\theta=148^{\circ}$).  All of the spectra except that of
36~Oph~A were taken using the Goddard High Resolution Spectrograph (GHRS)
instrument.  Of these spectra, all but that of Sirius utilized the high
resolution Echelle-A grating.  The Sirius spectrum was taken using the lower
resolution G140M grating, explaining the larger binsize of those data.  The
36~Oph~A data were obtained by the Space Telescope Imaging Spectrograph
(STIS), which replaced GHRS in 1997.  The resolution of that spectrum, which
was observed with STIS's E140H grating, is comparable to that of the
Echelle-A GHRS spectra.

     The spectra in Figure 1 are plotted on a heliocentric velocity scale.
All show broad, saturated H~I absorption near line center, and narrower
deuterium (D~I) absorption about 80 km~s$^{-1}$ blueward of the H~I
absorption.  Also plotted are the assumed intrinsic stellar Lyman-$\alpha$
lines (solid lines) and the best estimates for the interstellar absorption
(dotted lines), which fit the data only for 31~Com and $\beta$~Cas.  These
LISM absorption estimates are based on previously published work,
which we now discuss briefly.

     As mentioned in \S 1, the $\alpha$~Cen data provided the first evidence
for heliospheric absorption.  Both members of the $\alpha$~Cen binary system
were observed.  The existence of two independent data sets for the same line
of sight allowed for a very nice consistency check, and \citet{jll96}
demonstrated that the results of their analysis were the same for both the
$\alpha$~Cen~A and $\alpha$~Cen~B data.  It is the $\alpha$~Cen~B spectrum
that we display in Figure 1, since it has somewhat higher signal-to-noise
(S/N) than the $\alpha$~Cen~A data.

     Using the D~I line to define the central velocity and temperature of the
interstellar material, \citet{jll96} found they could not fit the
$\alpha$~Cen H~I profile with LISM absorption alone.  Excess H~I absorption
existed on both sides of the Lyman-$\alpha$ line (see Fig.\ 1), especially on
the red side.  Heliospheric absorption is believed to be responsible for the
excess absorption on the red side, and stellar astrospheric absorption may be
responsible for the less prominent excess absorption on the blue side
\citep{kgg97}.

     The LISM H~I column density toward $\alpha$~Cen could not be constrained
well at all, so two models of the absorption were presented: one with a
deuterium-to-hydrogen (D/H) ratio similar to previously measured values for
the local interstellar cloud, ${\rm D/H}\approx 1.5\times 10^{-5}$
\citep{jll98}, and one with a much smaller D/H value
(${\rm D/H}=6\times 10^{-6}$).  In Figure 1, we show the model with the
generally accepted D/H value.

     The analysis of the 36~Oph data proved to be very similar to that of
$\alpha$~Cen, with the apparent existence of excess Lyman-$\alpha$ absorption
on both the red and blue sides of the line (see Fig.\ 1).  \citet*{bew00}
argue that only contributions of both the heliospheric and
astrospheric absorption can account for all of the excess absorption.

     The Sirius data were first presented by \citet{pb95a,pb95b}.
The Sirius Lyman-$\alpha$ spectrum has lower resolution and lower S/N than
the other spectra in Figure 1.  Furthermore, unlike the other lines of sight,
there are two interstellar components instead of just one.  Nevertheless,
\citet{pb95a,pb95b} found that if they constrained the H~I absorption by
assuming the LISM temperature and local D/H ratio reported by \citet{jll93}
toward Capella (T=7000 K and ${\rm D/H}=1.65\times 10^{-5}$,
respectively), they could not account for excess H~I absorption on both the
blue and red sides of the line (see Fig.\ 1).  They interpreted the larger
excess on the blue side to be due to an optically thick stellar wind, and the
red side excess to be due to absorption from an evaporative interface between
the local cloud and surrounding hot ISM.

     \citet*{vvi99} presented a different interpretation of these
data.  They proposed that the red excess is due to heliospheric absorption
and the blue excess is due to analogous astrospheric absorption.  They
supported this interpretation using a self-consistent kinetic/gasdynamic
model, which suggests that a combination of heliospheric and astrospheric
material can produce the required amount of absorption.  The estimated LISM
absorption shown in Figure 1 is from \citet*{vvi99}.

     The 31~Com, $\beta$~Cas, and $\epsilon$~Eri data shown in Figure 1
were first analyzed by \citet{ard97}.  The intrinsic stellar
Lyman-$\alpha$ lines we assume in the figure are somewhat different than
those derived by \citet{ard97}, but the LISM parameters used to
compute the interstellar absorption profiles are nearly identical.  The 31~Com
and $\beta$~Cas spectra can be fitted beautifully by single interstellar
absorption components, but the $\epsilon$~Eri spectrum is more complex.
The H~I absorption is significantly blueshifted relative to the D~I
absorption, indicating a substantial amount of excess H~I absorption on
the blue side of the line (see Fig.\ 1).  \citet{ard97} interpreted
this excess absorption to be astrospheric in origin.

\section{The Models}

     We construct many models of the heliosphere using the Boltzmann
technique described by \citet*{asl98}, \citet{hrm99}, and \citet*{hrm00},
and we also present a four-fluid model for comparison (see Table 1).
Lyman-$\alpha$ absorption is computed for each model for
comparison with the HST data.  The models assume standard values for the
fully ionized solar wind at 1 AU:  ${\rm n(H^{+})}=5.0$ cm$^{-3}$,
$v=400$ km~s$^{-1}$, and $T=10^{5}$ K.
For the inflowing partially ionized LISM, we assume a velocity of
26 km~s$^{-1}$ and a neutral hydrogen density of ${\rm n(H)}=0.14$ cm$^{-3}$,
which are consistent with observations \citep{jll00,rl95,eq94,gg93}.

     Other input parameters are varied, and the parameters assumed for the
various models are listed in Table 1.  Observations suggest a likely value
of ${\rm T}\approx 8000$ K for the interstellar temperature \citep{bew98,mw96}
and a value of ${\rm n(H^{+})}\approx 0.1$ cm$^{-3}$ for the interstellar
proton density \citep{bew97}, so these are the numbers assumed for most of our
models.  In Models 8 and 9, we also experiment with smaller values that are
still plausible assumptions for the real LISM (see Table 1).

     However, the parameter we vary the most is a parameter, $\alpha$,
relating the proton pressure to the total LISM plasma pressure,
$P=\alpha n(H^{+}) kT$.  For the known velocity of the inflowing LISM, the
Mach number M is related to $\alpha$ by ${\rm M}=220(\alpha T)^{-0.5}$,
with $T$ measured in K.  For a pure hydrogen plasma, $\alpha=2$, based on the
assumption that the electrons and the protons of the plasma have equal number
densities and temperatures.  The existence of significant cosmic ray and
magnetic field pressure in the LISM could potentially increase $\alpha$ well
above 2.  Neither of these additional sources of pressure is well constrained
by observation, so we experiment with a wide range of $\alpha$'s (see Table 1).

     In Figure 2, we display the temperature and density distributions for
protons and neutral hydrogen in the heliosphere for Model 1.  Panel 2a shows
the locations of the heliospheric boundaries, and streamlines in panel 2b 
indicate the paths of protons in the model.  The LISM plasma is heated at the
bow shock (BS) and diverted around the heliosphere (see Fig.\ 2a-b).  The
tangential discontinuity of the heliopause (HP) separates the LISM plasma
from the solar wind.  In Model 1, the distances from the Sun to the HP and to 
the BS in the upwind direction are 100 AU and 260 AU, respectively.  Inside
the heliosphere, the cool, fast solar wind goes through a termination shock
(TS), which is at 62 AU in Model 1.  The solar wind is heated to temperatures
in excess of $10^6$ K and diverted tailward.  The neutral temperatures and
densities are computed by taking moments of the particle distributions
computed by the Boltzmann code \citep*[see][]{hrm00}.  In the region
between HP and BS, the overdensity in neutral hydrogen referred to as the
hydrogen wall is clearly visible in Figure 2d.  Downwind of this hydrogen
wall, neutral hydrogen is depleted but hot ($\approx 10^5$ K, see Fig.\ 2c).
Note that if $\alpha > 6$ then ${\rm M} < 1$ (assuming ${\rm T}=8000$ K), in
which case there will be no bow shock (Models 4--7).  A more in-depth
description of Models 1--10 and their properties will be provided in a future
paper (M\"uller, Zank, \& Wood 2000, in preparation).

     The LISM protons are shock heated at the bow shock, and if there is no
bow shock they are still heated by adiabatic compression at the heliopause.
The interstellar neutrals are then heated by charge exchange with these
protons, creating the high temperatures seen in Figure 2c.  Charge exchange
with outflowing solar wind protons also occurs, resulting in a flux of hot
neutrals back through the heliopause and bow shock.  Further charge exchanges
can deposit additional energy both in the hydrogen wall and in front of the
bow shock, and such processes can have measurable effects on the heliospheric
structure \citep{gpz96}.  In general, higher heliospheric hydrogen
temperatures and densities are observed for models with lower values of
$\alpha$, which is not surprising since a lower $\alpha$ corresponds with a
higher Mach number for the inflowing LISM and therefore more heating.
Because higher temperatures and column densities will produce broader
Lyman-$\alpha$ absorption profiles, models with higher Mach numbers will
also generally produce more absorption.

     The Boltzmann models provide us with velocity distributions throughout
the heliosphere that we can use to compute the absorption profiles.  In
Figure 3, radial velocity distributions are displayed for neutral hydrogen
particles in the upwind and downwind directions for Model 1.  In order to
create these distributions, particles are summed within $100\times100$ AU
boxes centered on the locations indicated in the figure.  Poissonian error
bars ($N^{0.5}$) are displayed for each velocity bin.  The bin size of the
histogram is 6 km~s$^{-1}$.  Gaussians have been fitted to the distributions
(dashed lines), and the poor fits illustrate the non-Maxwellian
character of the distributions.

     The complexity of the distributions is due in part to the fact that the
neutral particles are created by charge exchange with protons throughout the
heliosphere with very different temperatures and flow velocities.  The main
peak consists mostly of neutral particles created by charge exchange with
heated LISM protons in the hydrogen wall.  Following the nomenclature used
in \citet{gpz96}, these are called ``component 1'' neutrals.  The
distributions also contain some particles created by charge exchange with the
very hot, decelerated solar wind protons in between the termination shock and
heliopause, which for the upwind case in Figure 3 are seen after flowing back
across the heliopause.  These particles, the ``component 2'' neutrals, are
particularly prevalent in the far wings of the distributions.  The
component 2 neutrals are far more abundant in downwind directions, explaining
why the downwind distribution is broader and has more extended wings than the
upwind distribution.  Finally, there are a few particles in the distributions
created by charge exchange with solar wind protons near the Sun, thereby
creating a small population of neutrals with speeds of about
$+400$ km~s$^{-1}$ (see Fig.\ 3), which are referred to as ``component 3''
neutrals.  At the upwind location shown in Figure 3, the peak
of the distribution lies at about $-16$ km~s$^{-1}$, which represents a
significant deceleration from the $-26$ km~s$^{-1}$ speed of the undisturbed
LISM.  At the downwind location shown in the figure, the average speed is about
$+28$ km~s$^{-1}$, slightly faster than the $+26$ km~s$^{-1}$ LISM flow speed.

     In Figure 4, we show the H~I Lyman-$\alpha$ absorption predicted by
Model 1 for an upwind and downwind line of sight.  In computing such
absorption profiles, we first divide the radial path through the heliosphere
into many segments.  We assign an H~I density to each segment based on the
number of particles found within a $4\times5$ AU box containing the segment.
To get viable statistics for the distribution of radial velocities for the
segment, we must use a larger area of influence to determine the distribution
histogram.  We use a circle with a radius of 30 AU containing the
segment for this purpose.

     Velocities can be linearly mapped onto wavelengths using the familiar
relation
\begin{equation}
\lambda=\lambda_{0}\left( 1-\frac{v}{c} \right)^{-1},
\end{equation}
where $\lambda_{0}=1215.6701$~\AA\ is the rest wavelength of Lyman-$\alpha$.
Thus, the velocity distribution defines a line profile function,
$\phi_{\lambda}$, which is normalized so that
\begin{equation}
\int^{\infty}_{0} \phi_{\lambda} d\lambda = 1.
\end{equation}
The opacity profile in frequency space is (in cgs units)
\begin{equation}
\tau_{\nu}=0.02654 f N \phi_{\nu},
\end{equation}
where $f$ is the oscillator absorption strength and $N$ is the column density
(i.e., the density times the pathlength of the line segment in question).
For Lyman-$\alpha$, $f=0.4164$ \citep{dcm91}.  Since
$\phi_{\nu}=(\lambda^{2}/c) \phi_{\lambda}$, the opacity profile in
wavelength space is
\begin{equation}
\tau_{\lambda}=\frac{0.02654 f N \lambda^{2}}{c} \phi_{\lambda}.
\end{equation}
We compute these opacity profiles for each segment of the line of sight and
then add them all up.  The absorption profile is then simply
$I_{\lambda}=I_{0}\exp(-\tau_{\lambda})$, where $I_{0}$ is the assumed
background Lyman-$\alpha$ flux.  In Figure 4, we simply assume $I_{0}=1$
for all wavelengths.

     In the procedure described above, we have assumed that the line profile
is regulated solely by Doppler motions.  This is a very good approximation
here since the velocity distributions are much broader than the core of the
natural line broadening profile, and column densities are too low for the
Lorentzian wings of the natural profile to become evident.  We have also
ignored the fine structure of Lyman-$\alpha$, but this has only a very
minor effect since the two fine structure components are separated by
only 1.3 km~s$^{-1}$.

     The H~I properties displayed in Figures 2c-d were computed by taking
moments of the velocity distribution functions.  In Figure 4, we show
absorption profiles computed directly from these moments (dashed lines).
These profiles, however, are significantly different from those computed
directly from the distributions, indicating that flow velocities and
temperatures computed from moments do not define the characteristics of the
distribution well enough to be used to accurately determine absorption
profiles.  Accurate absorption profiles can be computed only from the
distributions themselves.

\section{Comparing the Models with the Data}

     In Figures 5 and 6 we show the heliospheric absorption predicted for the
lines of sight observed by HST for Models 1--7.  The heliospheric absorption
is combined with the LISM absorption (dotted lines) and then convolved with
the instrumental line spread function before being plotted.  For the sake of
clarity, we display four of the models in Figure 5 and four in Figure 6,
with Model 4 being shown in both.  We have zoomed in on the red side of the
absorption profiles since that is where most of the heliospheric absorption
is located, astrospheric absorption being the more likely explanation for
any excess absorption on the blue side (see \S 2).

     The models certainly have no trouble producing observable absorption.
In fact, the models tend to predict too much absorption, especially downwind.
This downwind discrepancy may be even more dramatic than the figures suggest,
because our models only extend 1000 AU in that direction, which may
be far enough for lines of sight with $\theta \leq 125^{\circ}$ but
is probably not far enough to incorporate all the heliospheric H~I
further downwind.  Thus, we will underestimate the amount of
heliospheric absorption toward Sirius and $\epsilon$~Eri.

     The absorption exhibits some interesting nonlinear behavior, which is
particularly pronounced in the downwind direction.  As expected, the amount
of absorption decreases with increasing $\alpha$ between $\alpha=2$ and
$\alpha=5$, but the absorption {\em increases} between $\alpha=5$ and
$\alpha=9.6$ before decreasing once again between $\alpha=9.6$
and $\alpha=18$ (see, e.g., the $\epsilon$~Eri panels of Figs.\ 5-6).
Similar behavior is observed upwind, although it is not as dramatic.  The
transition between decreasing absorption and increasing absorption at
$\alpha=5$ is presumably associated with the boundary between a
supersonic and a subsonic LISM wind, which is at $\alpha=6$.

     Unfortunately, none of the models does a very good job in reproducing
the observations.  Some of the apparent disagreement is illusory, however,
because it can be fixed by tweaking either the assumed stellar Lyman-$\alpha$
profile or the assumed LISM absorption, or both.

     In Figure 7, we show how this is done for two examples: the $\alpha=2.0$
model for the 36~Oph line of sight and the $\alpha=12.5$ model for the 31~Com
line of sight.  In both cases, we allow the assumed LISM absorption
parameters to be altered within the uncertainties derived in the initial
analyses of these data, which are determined partially from the analyses of
other LISM lines \citep*{ard97,bew00}.  We also alter the
shapes of the assumed stellar profiles to try to make the models fit the
data (solid lines: original, dashed lines: altered profiles).

     However, there are limits to the shapes a reasonable
profile can have.  This is especially true for 31~Com.  The 31~Com
Lyman-$\alpha$ emission line is very broad, meaning we cannot allow the
slope of the line to become very large within the absorption region.
This limits how effective profile changes are in helping us to force the
heliospheric absorption model to fit the data.  The narrower 36~Oph emission
line allows us more leeway in altering the assumed profile.  Thus, the
changes from the original profiles to the new profiles
in Figure 7 are much larger for 36~Oph than for 31~Com.

     Another problem for 31~Com is that there is no evidence for excess
absorption on the blue side of the line that would suggest the presence of
astrospheric absorption, meaning that we cannot completely ignore the quality
of fit on that side of the line, in contrast with the 36~Oph example.
The quality of the 31~Com fit in Figure 7 is noticeably worse than
that of the simpler LISM-only fit in Figure 1.  Nevertheless, we conclude
that the 31~Com fit in Figure 7 is just good enough to claim that the
$\alpha=12.5$ model (Model 6) is barely consistent with the 31~Com
observations at $\theta=73^{\circ}$.  Based on Figure 7, we also claim that
the $\alpha=2.0$ model (Model 1) is consistent with the 36~Oph observations
at $\theta=12^{\circ}$.

     We have performed an analysis like that in Figure 7 for all combinations
of models and data, and in Figure 8 we show the best fits that can be
obtained by such alterations of stellar profile and LISM absorption for
Models 1--7.  Many of the models that did not appear to agree well with
the data in Figures 5--6 do fit the data well in Figure 8, but many still do
not.  In general, discrepancies near the base of the
absorption (e.g., at $15<v<25$ km~s$^{-1}$ for $\alpha$~Cen) are much harder
to remedy than discrepancies farther in the wing of the absorption
(e.g., at $v>25$ km~s$^{-1}$ for $\alpha$~Cen).  Large discrepancies near
the base can generally only be fixed by introducing unreasonable fine
structure into the assumed stellar Lyman-$\alpha$ profile.

     Based on the results displayed in Figure 8, we provide in Table 1 our
evaluation of which observed lines of sight are inconsistent with which
models.  None of the models discussed to this point (Models 1--7) are
consistent with every line of sight.  The low $\alpha$ models do better
upwind and the high $\alpha$ models do better downwind.  In general, the
problem with the models that do not fit the data is that they
predict too much absorption, the exception being the 36~Oph line of sight
for which most of the models predict too little absorption.  Model 7 is
the only model that does not predict way too much absorption along
the downwind line of sight to $\epsilon$~Eri.

     For Models 8 and 9 in Table 1, we repeated the $\alpha=2.0$ model but
with a lower LISM proton density and temperature, respectively, in order to
see what effect this has on the predicted absorption (see \S 3).  The
n(H$^{+}$) and T values assumed for these models are not the most likely
values based on our current understanding of the LISM, but are still within
the realm of possibility (see Table 1).  Figure 9 compares the absorption
predicted by all the $\alpha=2.0$ models.  The figure shows that lowering the
proton density decreases the amount of absorption downwind but has little
effect elsewhere.  Decreasing the temperature decreases the amount of
absorption in all directions.

     The changes in absorption indicated in Figure 9 are fairly modest,
however, and certainly not enough to change the results shown in Figure 8,
which suggest that $\alpha=2.0$ models are consistent only with the 36~Oph
line of sight.  If lower proton densities and/or temperatures were assumed
for the models with other assumed $\alpha$'s, it might allow some of them to
become consistent with some of the observed lines of sight.  The
$\alpha=12.5$ model could be made consistent with the
$\epsilon$~Eri data, for example (see Fig.\ 8), but it would worsen the
disagreement with the 36~Oph data.

     Thus, we conclude that small changes in n(H$^{+}$) and T will not solve
the overall disagreement between the current Boltzmann models and the data.
The models predict too much absorption downwind and sidewind relative to the
amount of absorption observed upwind.  Given the non-linear response of
heliospheric models to different boundary conditions, we cannot completely
exclude better agreement between data and models in another area of parameter
space, perhaps one where the solar wind parameters are different than
those that we assume.  Moreover, barring unaccounted systematic errors in the
assumed stellar profiles and the ISM absorption component, the heliospheric
models employed in this study do not yet include all possibilities of
interaction in the heliosphere.  All of the models neglect the magnetic
fields of both the solar wind and interstellar medium, the latter being very
poorly constrained observationally.  The Boltzmann kinetic models neglect
neutral-neutral collisions, as well as proton-neutral collisions that are not
accompanied by charge exchange, and neutral depletion due to photoionization
close to the Sun.

     Ironically, the one model listed in Table 1 that is consistent with all
lines of sight is the four-fluid model (Model 10), which uses a less
sophisticated treatment of the heliospheric H~I than the Boltzmann models.
Figure 10 shows the comparison of predicted and observed absorption for
Model 10, both before and after the assumed stellar profile and ISM
parameters are tweaked to maximize the quality of fit (as was done in Fig.\ 8
for Models 1--7).  The agreement is good enough to claim that the model is
consistent with all six lines of sight.  Note that the absorption predicted
by the four-fluid Model 10 is vastly different from that predicted by the
Boltzmann Model 1 (see Fig.\ 10 and $\alpha=2.0$ model in Fig.\ 5), despite
the fact that their input parameters are identical (see Table 1).  This
illustrates just how sensitive the predicted absorption is to how the H~I
velocity distributions are treated in the models.

     There are several ways to interpret the results illustrated in Figure 10.
It should be understood that the four-fluid model and the Boltzmann/kinetic
models represent somewhat different physics for the neutral H distributions.
The underlying assumption inherent in the four-fluid model is that some
scattering of neutral H is present to ensure that the distributions of the
individual components, {\em if not the full distribution}, is approximately
Maxwellian.  This process is absent in the kinetic models (see Zank et al.\
1996 and Williams et al.\ 1997 for discussions concerning the four-fluid
model assumptions).  Figure 10 indicates that a good fit to the
Lyman-$\alpha$ data along the six sightlines requires that the heliospheric
neutral H distribution be very like that produced by the four-fluid model.
That Boltzmann Model 1 and the four-fluid Model 10 produce quite different
absorption profiles suggests the possibility that additional physical
processes with respect to H may need to be incorporated in the Boltzmann
codes.  Unfortunately, we have not yet explored parameter space in sufficient
detail to rule out the possibility that a Boltzmann model does in fact exist
that results in acceptable agreement with the Lyman-$\alpha$ data, assuming
reasonable input parameters.  In addition, other factors not included,
explicitly or implicitly, in either the Boltzmann or four-fluid models (such
as magnetic fields) may still be required to fully address these issues.

\section{Summary}

     We have compared H~I Lyman-$\alpha$ absorption profiles computed using
fully self-consistent kinetic/hydrodynamic models of the heliosphere with
profiles observed in UV spectra obtained by HST.  We make this comparison
for six different lines of sight through the heliosphere toward six nearby
stars.  Our results are summarized as follows:
\begin{description}
\item[1.] Most of our models use a Boltzmann particle code for the neutrals,
  allowing us to estimate neutral velocity distributions throughout the
  heliosphere.  These velocity distributions are far from Maxwellian, having
  extended wings that lead to broad Lyman-$\alpha$ absorption profiles.
\item[2.] Average temperatures and flow velocities can be computed by taking
  moments of the velocity distributions, and Lyman-$\alpha$ absorption
  profiles can be computed from these moments.  However, we find that
  such profiles do not generally agree well with more accurate profiles
  computed directly from the distributions.  Thus, the moments of the
  distributions apparently do not provide enough information about their
  shapes to yield accurate absorption profiles.
\item[3.] The amount of absorption predicted by the models generally
  increases as the assumed Mach number of the interstellar wind is increased
  (i.e., the $\alpha$ parameter is decreased).  However, there are
  some interesting non-linear effects near the transition between subsonic
  and supersonic models (i.e., near $M=1$).
\item[4.] The Boltzmann models tend to predict too much absorption in
  sidewind and downwind directions, and too little absorption upwind.  This
  problem is exacerbated at higher Mach numbers.  Varying the assumed
  interstellar temperature and proton density does not seem to help.
\item[5.] In contrast to the Boltzmann models, a four-fluid model that uses
  a less sophisticated, multi-fluid treatment of the neutrals is consistent
  with the data.  The absorption predicted by this model is very different
  from that predicted by a Boltzmann model with the same input parameters,
  emphasizing how crucial the treatment of the neutrals is for accurately
  predicting Lyman-$\alpha$ absorption profiles.
\item[6.] Possible reasons for the lack of success for the Boltzmann models
  in reproducing the data include the neglect of collisions not involving
  charge exchange in those models, or perhaps an insufficient exploration of
  parameter space.  There may also be other physical processes that have not
  been considered in either the Boltzmann or four-fluid models which could be
  important and affect our results, such as processes involving magnetic
  fields.
\end{description}

\acknowledgments

Support for this work was provided by NASA grant NAG5-9041 to the University
of Colorado. GPZ and HRM acknowledge the partial support of NASA grant
NAG5-6469.

\clearpage

\begin{deluxetable}{cccccccccccc}
\tablecaption{Model Parameters}
\tablecolumns{12}
\tablewidth{0pt}
\tablehead{
  \colhead{Model \#} & \colhead{Code} & \colhead{n(H$^{+}$)} &
    \colhead{T} & \colhead{$\alpha$} & \colhead{M} &
    \multicolumn{6}{c}{Consistent with Data?} \\
  \cline{7-12} \\
 \colhead{} & \colhead{} & \colhead{(cm$^{-3}$)} & \colhead{(K)} &
    \colhead{} & \colhead{} & \colhead{$\theta=12^{\circ}$} &
    \colhead{$52^{\circ}$} & \colhead{$73^{\circ}$} & \colhead{$112^{\circ}$}
    & \colhead{$139^{\circ}$} & \colhead{$148^{\circ}$}}
\startdata
1 & Boltzmann & 0.10 & 8000 & 2.0 & 1.7 & Y & N & N & N & N & N \\
2 & Boltzmann & 0.10 & 8000 & 3.5 & 1.3 & Y & N & N & N & N & N \\
3 & Boltzmann & 0.10 & 8000 & 5.0 & 1.1 & N & Y & Y & N & N & N \\
4 & Boltzmann & 0.10 & 8000 & 7.6 & 0.9 & N & Y & Y & N & N & N \\
5 & Boltzmann & 0.10 & 8000 & 9.6 & 0.8 & N & Y & N & N & N & N \\
6 & Boltzmann & 0.10 & 8000 &12.5 & 0.7 & N & Y & Y & Y & Y & N \\
7 & Boltzmann & 0.10 & 8000 &18.0 & 0.6 & N & Y & Y & Y & Y & Y \\
8 & Boltzmann & 0.05 & 8000 & 2.0 & 1.7 & Y & N & N & N & N & N \\
9 & Boltzmann & 0.10 & 6000 & 2.0 & 2.0 & Y & N & N & N & N & N \\
10& Four-fluid& 0.10 & 8000 & 2.0 & 1.7 & Y & Y & Y & Y & Y & Y \\
\enddata
\end{deluxetable}

\clearpage

\begin{figure}
\plotfiddle{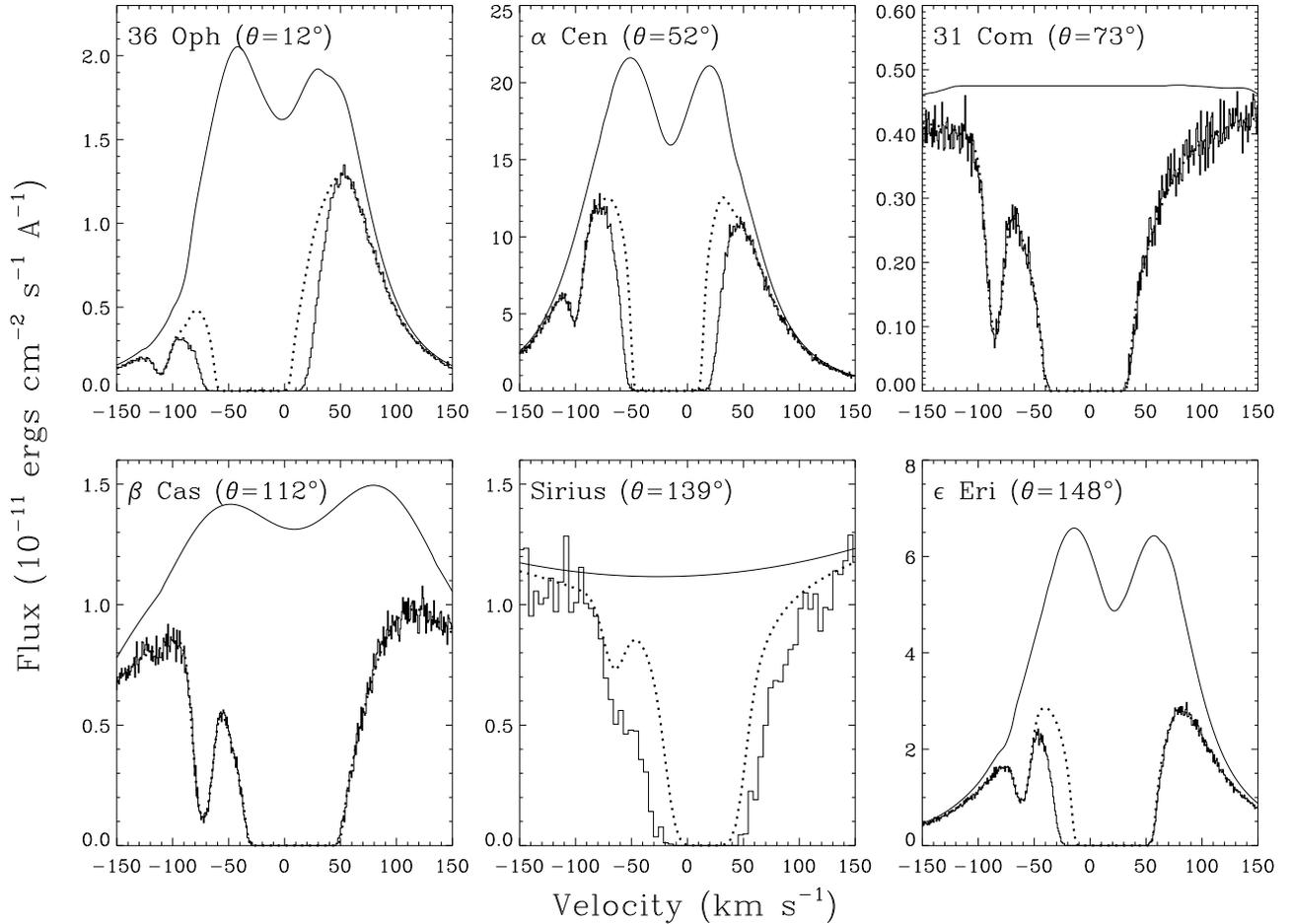}{3.5in}{90}{75}{75}{300}{0}
\caption{HST Lyman-$\alpha$ spectra of six stars, with the six lines of
  sight sampling different angles ($\theta$) relative to the upwind direction
  of the interstellar flow into the heliosphere.  Very broad H~I absorption
  is easily apparent in all the data, as is much narrower deuterium
  absorption about $-80$ km~s$^{-1}$ from the H~I absorption.  Each panel
  shows the assumed stellar profile (solid line) and interstellar absorption
  (dotted line), as determined from previously published work (see text).  In
  many cases, the interstellar absorption does not account for all of the
  observed absorption, presumably indicating the presence of heliospheric
  and/or astrospheric absorption.}
\end{figure}

\begin{figure}
\plotfiddle{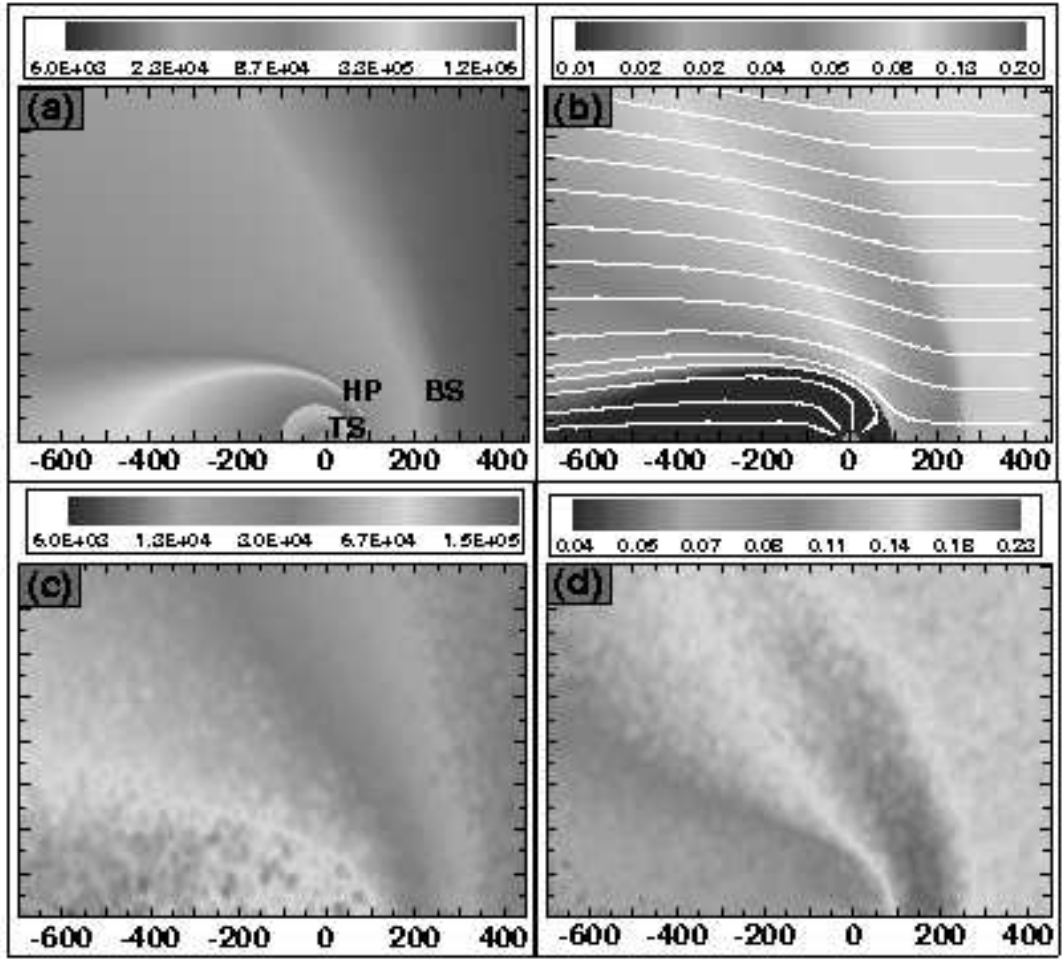}{4.5in}{0}{100}{100}{-235}{-190}
\caption{(a) Proton temperature, (b) proton density, (c) neutral hydrogen
  temperature, and (d) neutral hydrogen density distributions for Model 1.
  The positions of the termination shock (TS), heliopause (HP), and bow shock
  (BS) are indicated in (a), and streamlines indicating the plasma flow
  direction are shown in (b).  The distance scale is in AU.}
\end{figure}

\begin{figure}
\plotfiddle{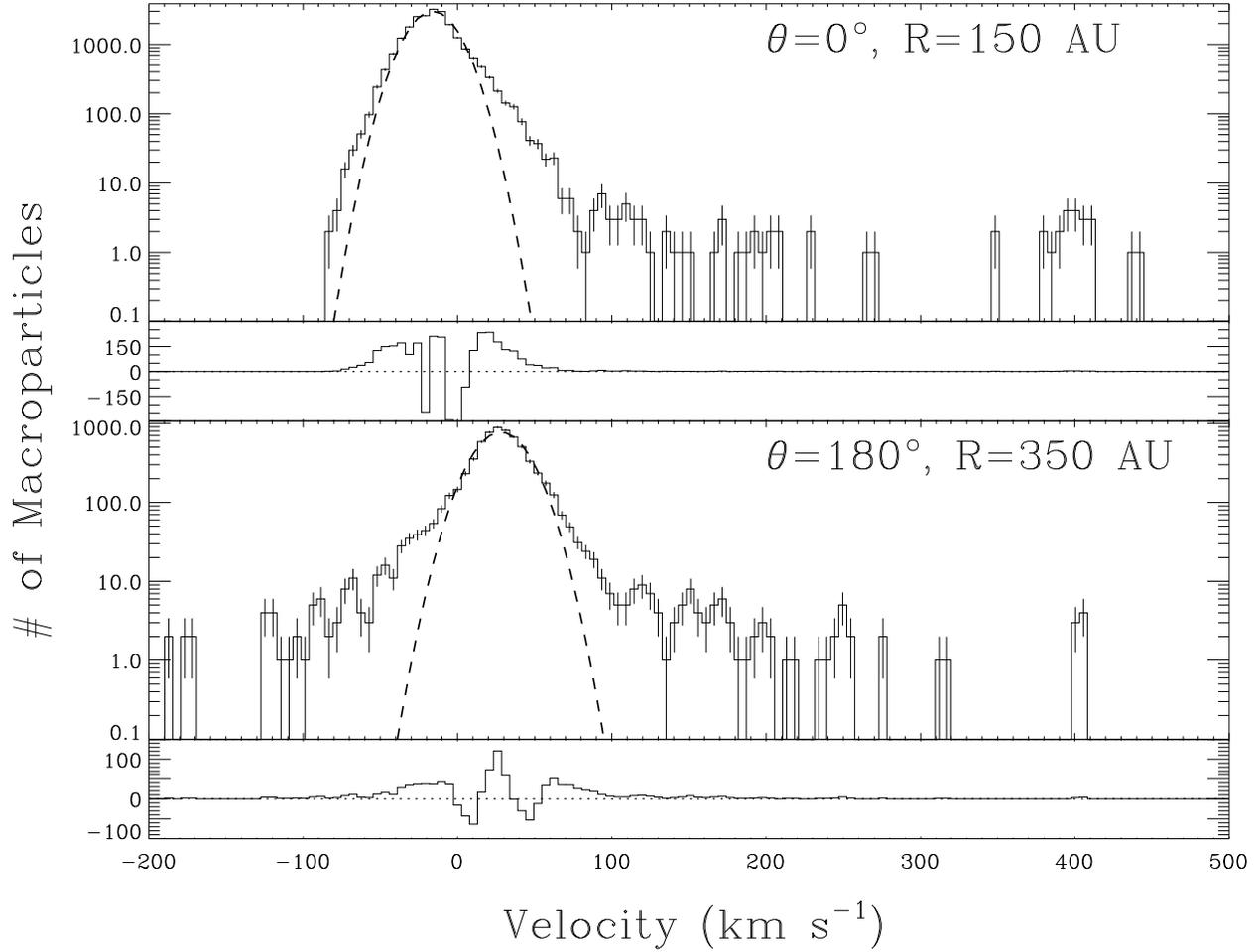}{3.5in}{90}{75}{75}{290}{0}
\caption{H~I radial velocity distributions for two different heliospheric
  locations based on Model 1, where $\theta$ is the angle
  relative to the upwind direction of the interstellar flow into the
  heliosphere, and R is the distance from the Sun.  The dashed lines are
  Gaussian fits to the distributions, and residuals of the fits are shown
  below each panel.  The poor quality of the fits illustrates the
  non-Maxwellian character of the distributions.}
\end{figure}

\begin{figure}
\plotfiddle{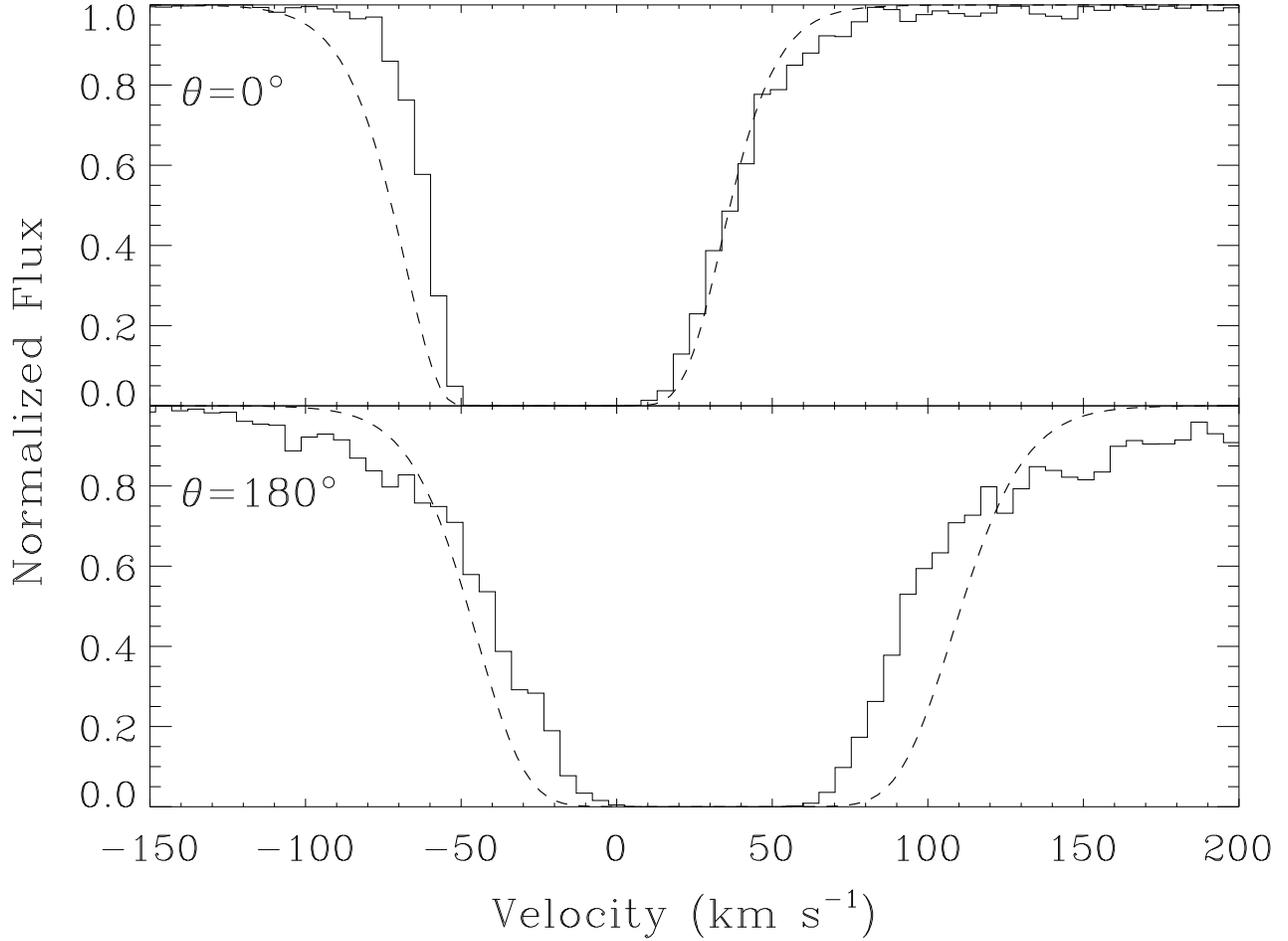}{3.5in}{90}{75}{75}{290}{0}
\caption{H~I Lyman-$\alpha$ absorption profiles for upwind
  ($\theta=0^{\circ}$) and downwind ($\theta=180^{\circ}$) lines of sight,
  computed for Model 1.  The dashed lines are profiles
  computed from temperatures and flow velocities calculated by taking
  moments of the velocity distributions, while the histograms are profiles
  computed directly from the distributions themselves.  The disagreement
  between the two demonstrates the importance of calculating absorption
  profiles directly from the distributions.}
\end{figure}

\begin{figure}
\plotfiddle{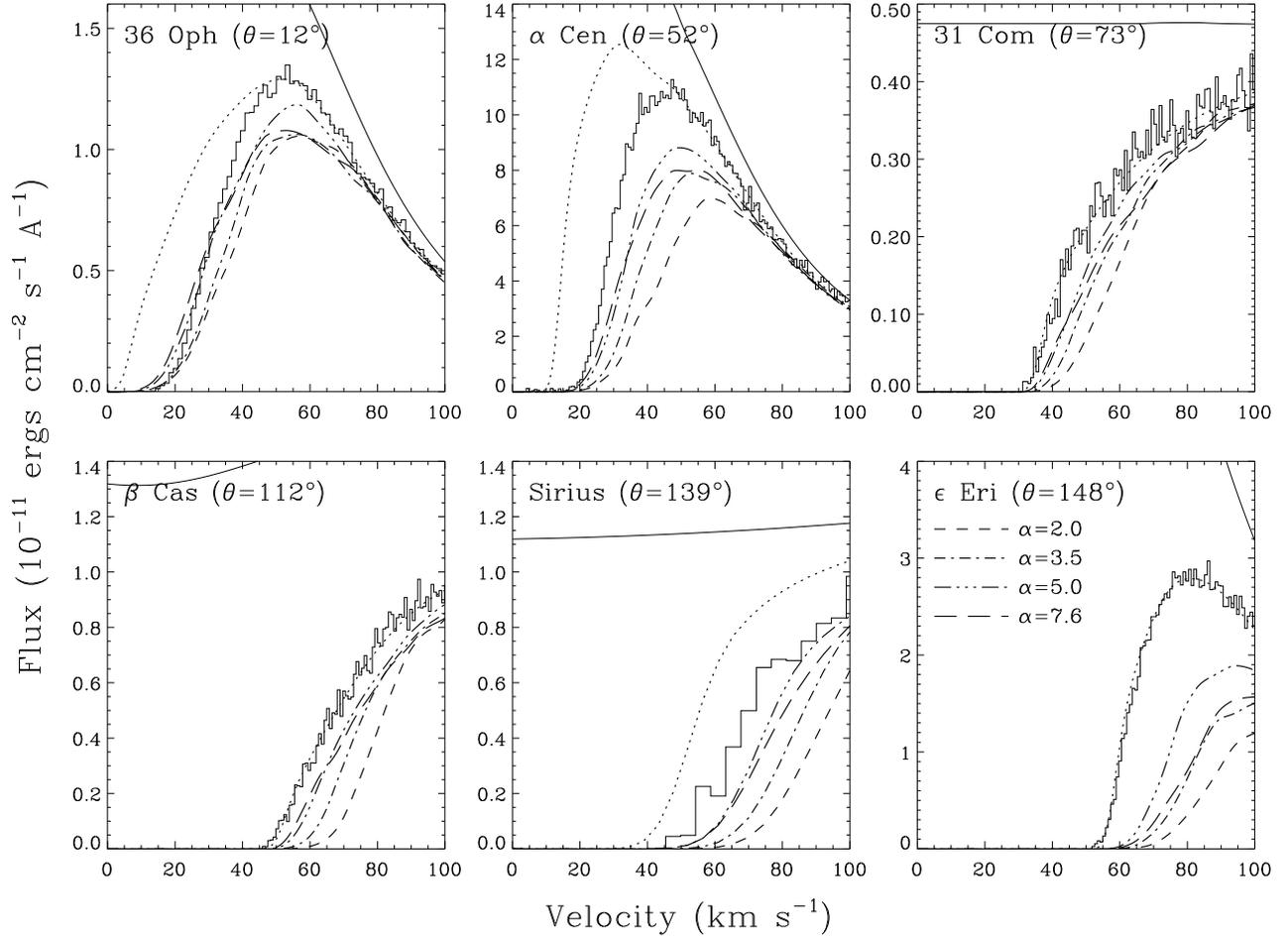}{3.5in}{90}{75}{75}{300}{0}
\caption{A reproduction of Fig.\ 1, zoomed in on the red side of the
  H~I absorption line, where we also show the absorption predicted by
  Models 1--4, which assume $\alpha=2.0-7.6$.}
\end{figure}

\begin{figure}
\plotfiddle{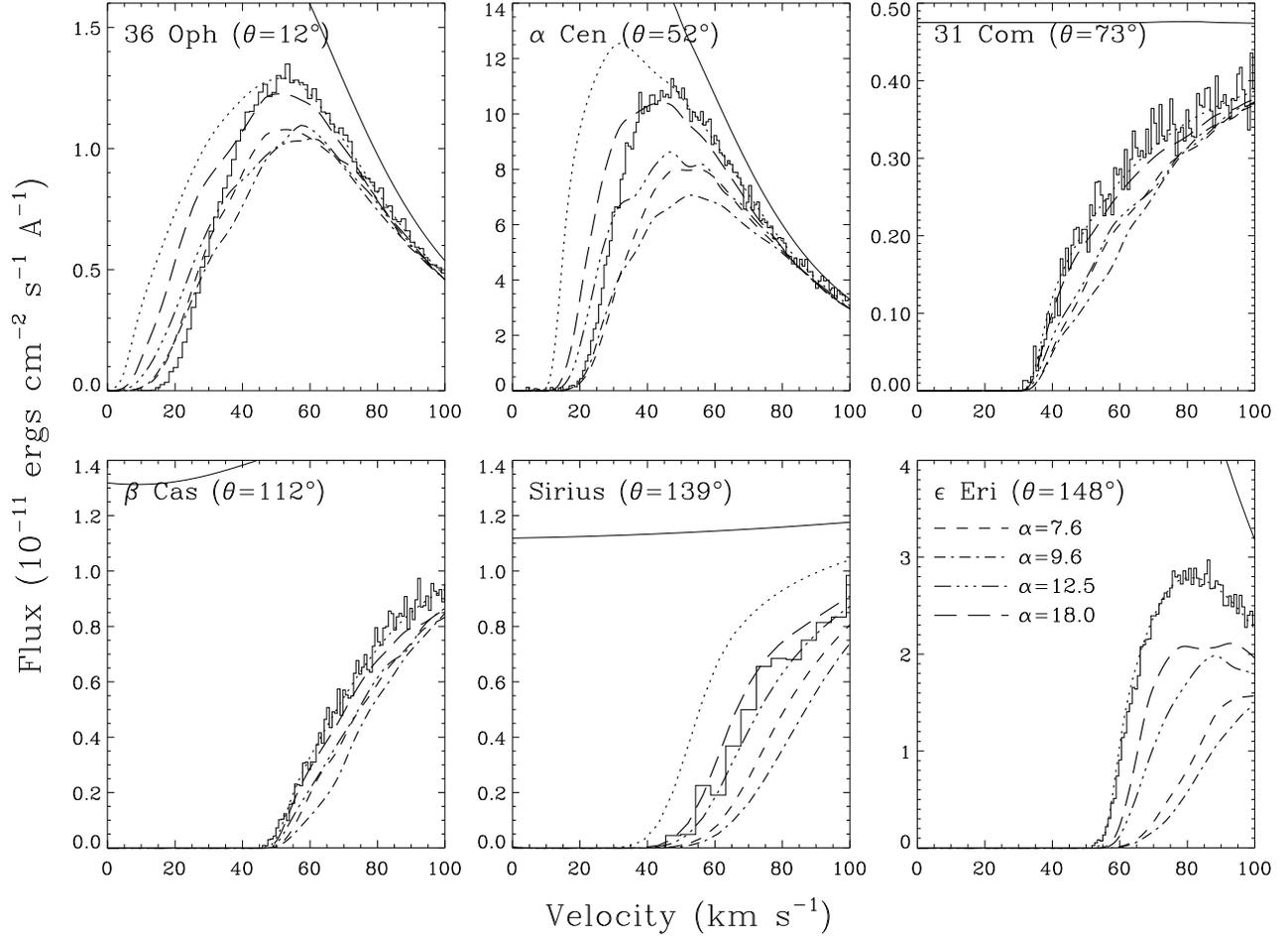}{3.5in}{90}{75}{75}{300}{0}
\caption{Same as Fig.\ 5, but for subsonic Models 4--7, which assume
  $\alpha=7.6-18.0$.}
\end{figure}

\begin{figure}
\plotfiddle{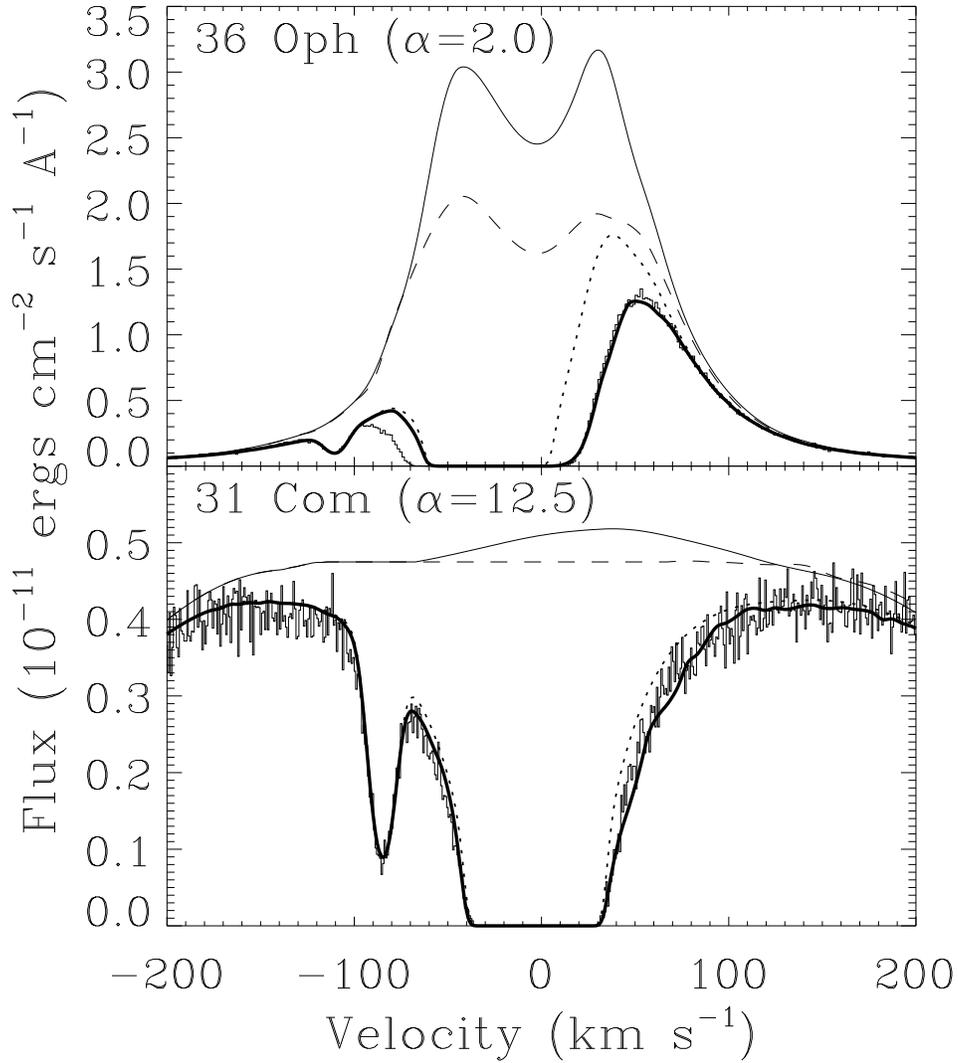}{4.5in}{0}{80}{80}{-270}{0}
\caption{Two examples of how models that do not appear to fit the data
  well in Figs.\ 5--6 can be made to fit the data by altering the assumed
  stellar Lyman-$\alpha$ profile and/or tweaking the assumed interstellar
  absorption parameters within the uncertainties determined from previously
  published analyses.  The dashed lines are the original stellar profiles,
  and the thin solid lines are the new ones.  In each panel, the dotted line
  represents the assumed interstellar absorption, and the thick solid line is
  the combination of the interstellar and heliospheric absorption for the
  model indicated in the figure, which in both cases fits the data reasonably
  well.}
\end{figure}

\begin{figure}
\plotfiddle{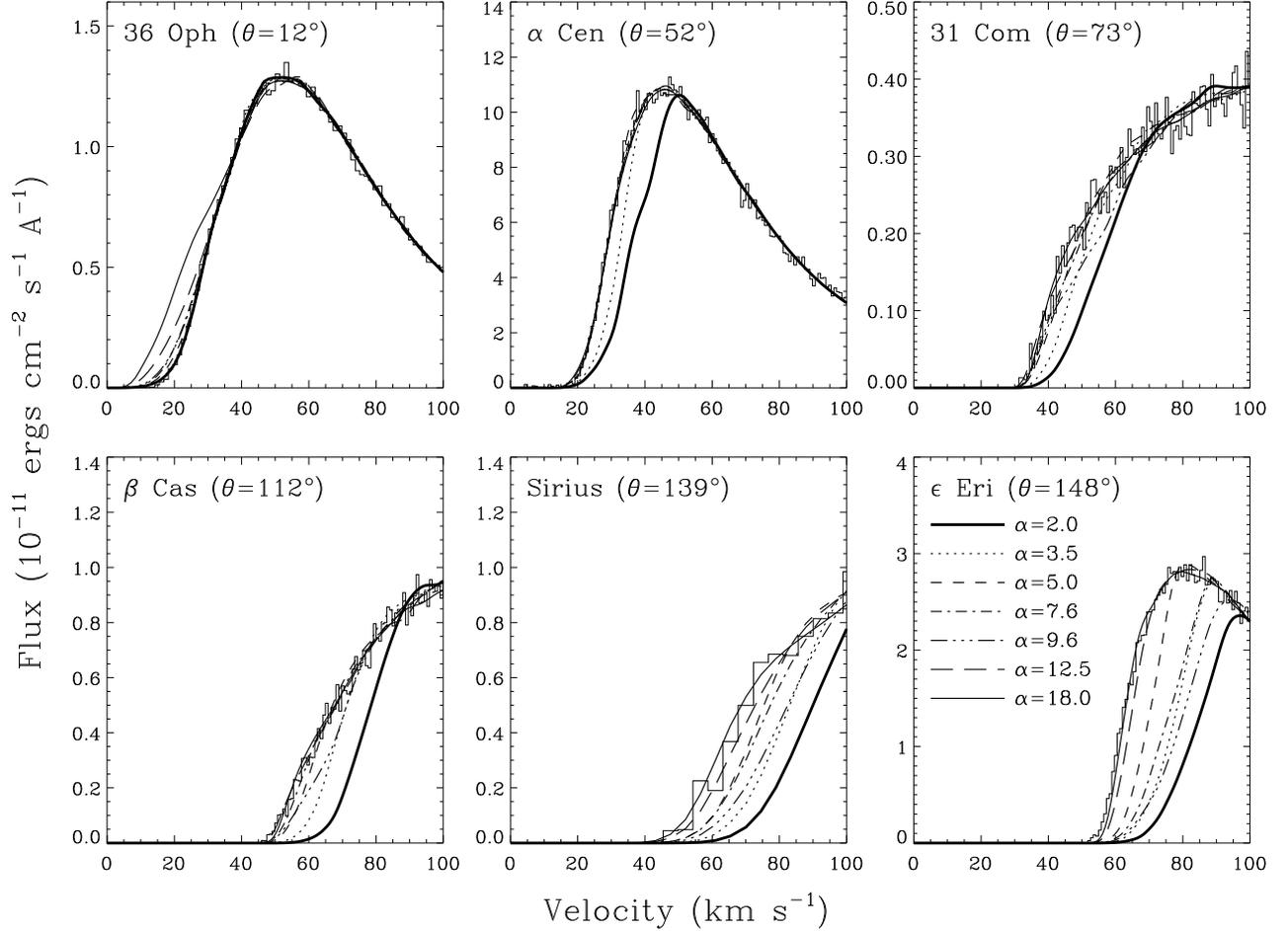}{3.5in}{90}{75}{75}{300}{0}
\caption{As suggested by Fig.\ 7, agreement between the models and
  observations can be greatly improved by tweaking the assumed stellar
  Lyman-$\alpha$
  line and/or ISM absorption.  This figure shows the best possible fits that
  can be obtained after doing this for all the models (Models 1--7) and lines
  of sight shown in Figs.\ 5--6.  This procedure results in good agreement
  with the data in some instances, but substantial disagreement remains in
  most cases.}
\end{figure}

\begin{figure}
\plotfiddle{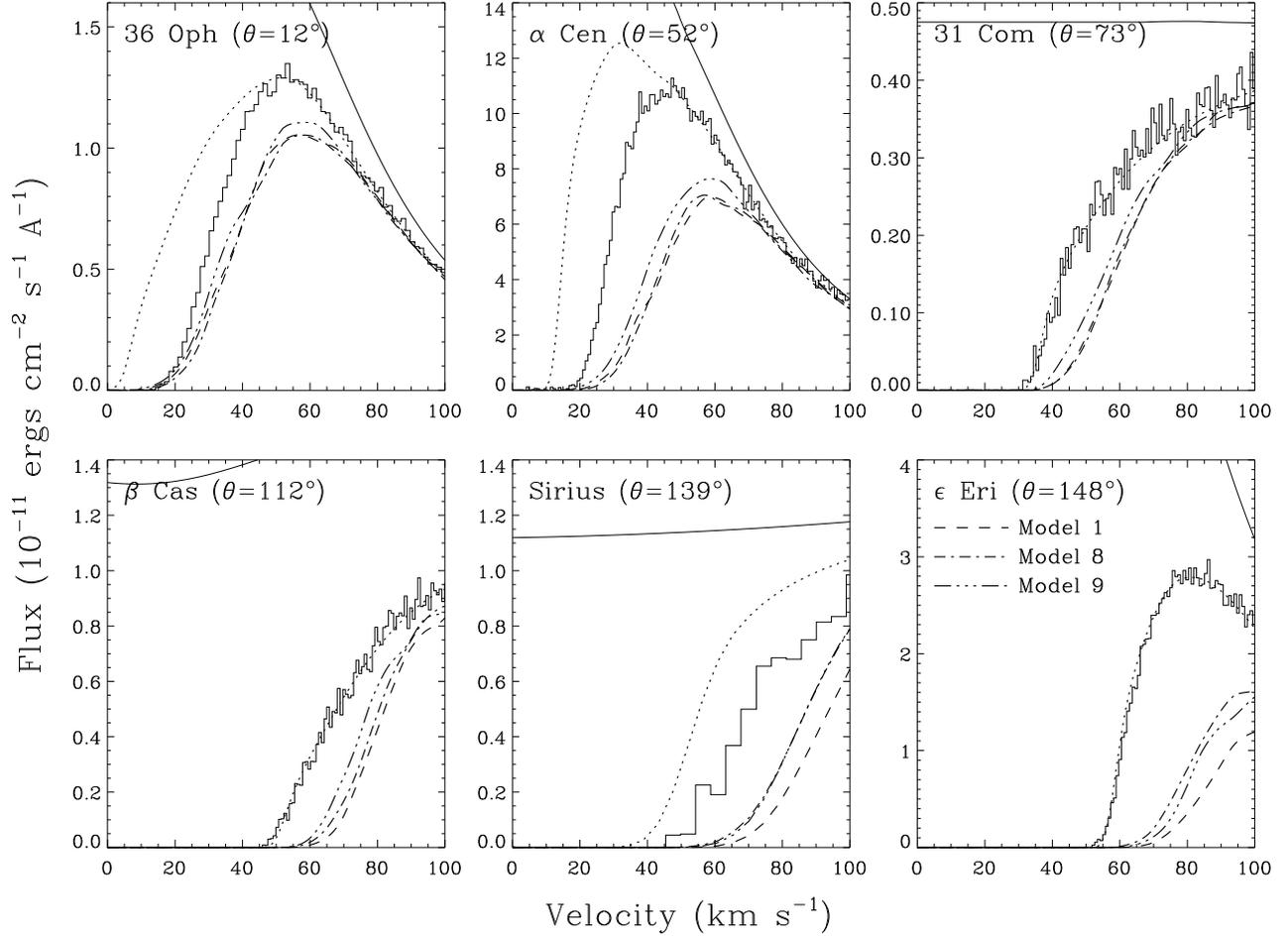}{3.5in}{90}{75}{75}{300}{0}
\caption{Comparison of the H~I absorption predicted by the three
  $\alpha=2.0$ Boltzmann models listed in Table 1, indicating the degree
  to which the absorption is decreased by reducing the assumed
  interstellar proton density (Model 8) and temperature (Model 9).}
\end{figure}

\begin{figure}
\plotfiddle{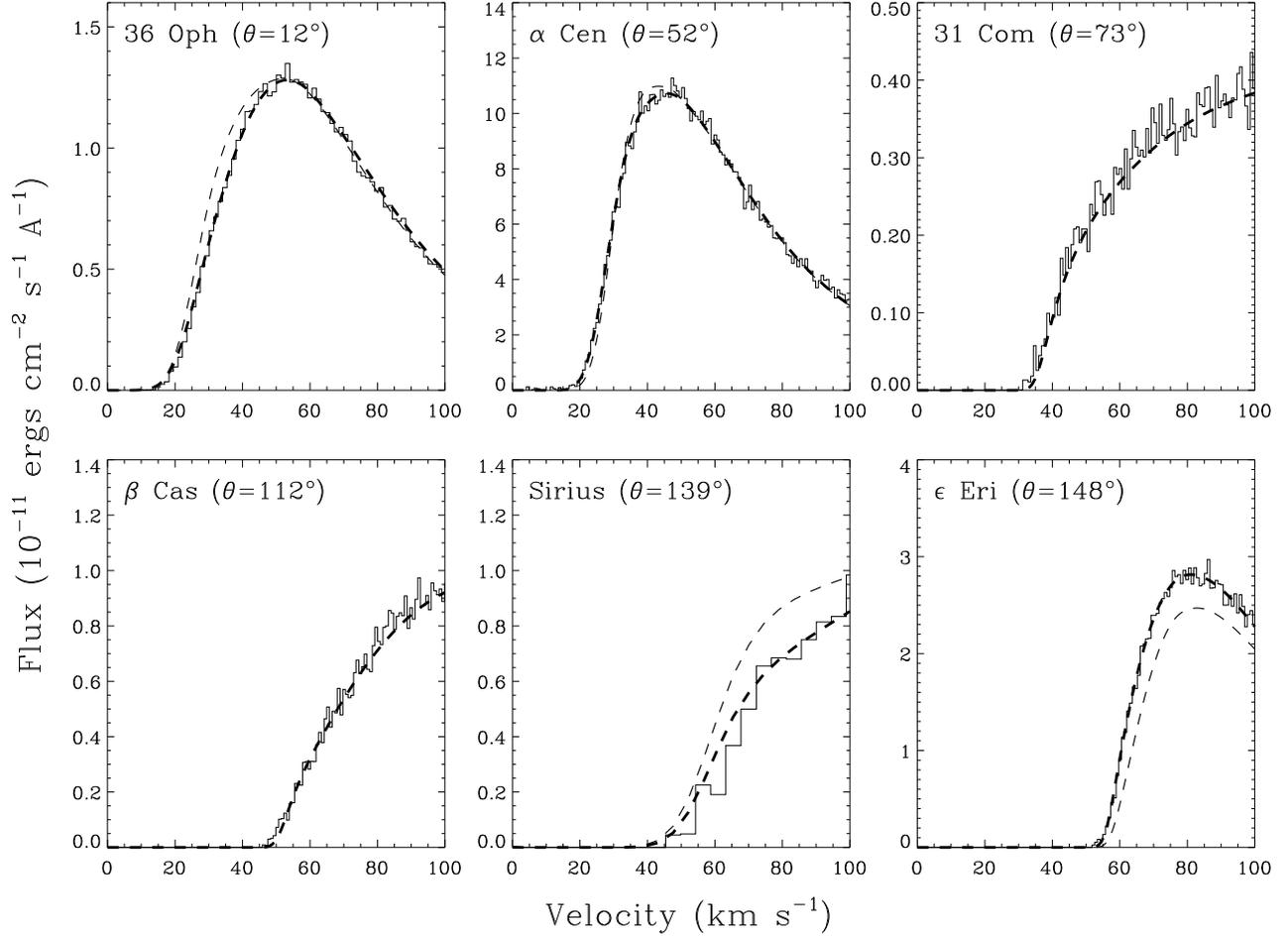}{3.5in}{90}{75}{75}{300}{0}
\caption{Comparison of the H~I absorption predicted by the $\alpha=2.0$
  four-fluid model listed in Table 1 (Model 10) and the observations, both
  before (thin dashed line) and after (thick dashed line) the assumed stellar
  line profile and ISM absorption are tweaked to maximize the agreement with
  the data.  Reasonably good agreement is observed for all six lines of
  sight.}
\end{figure}

\end{document}